%% file: main.tex
\documentclass[a4paper]{article}

\usepackage{INTERSPEECH2020}
\usepackage{amsmath,amsfonts,graphicx}
\usepackage{amssymb,textcomp,mathtools}
\usepackage{bm,upgreek,algorithm,hyperref}
\usepackage{multirow,booktabs,hhline,array}
\usepackage{cite,url,makecell,setspace,subcaption}

\pdfoutput=1

\def\thline{\noalign{\hrule height 1.0pt}}

\renewcommand{\vec}[1]{\bm{\mathrm{#1}}}

\title{Separating Varying Numbers of Sources with Auxiliary Autoencoding Loss}
\name{Yi~Luo, Nima~Mesgarani}
\address{
  Department of Electrical Engineering, Columbia University
}
\email{yl3364@columbia.edu, nima@ee.columbia.edu}

\begin{document}

\maketitle

\begin{abstract}
\input{abstract}
\end{abstract}

\section{Introduction}
\input{introduction}
\label{sec:introduction}

\section{Auxiliary Autoencoding Permutation Invariant Training}
\input{A2PIT}
\label{sec:A2PIT}

\section{Experiments}
\input{experiments}
\label{sec:exp}

\section{Conclusion}
\input{conclusion}
\label{sec:conclusion}

\section{Acknowledgments}
This work was funded by a grant from the National Institute of Health, NIDCD, DC014279; a National Science Foundation CAREER Award; and the Pew Charitable Trusts.

\bibliographystyle{IEEEtran}
\bibliography{refs}

\end{document}

%% file: abstract.tex
Many recent source separation systems are designed to separate a fixed number of sources out of a mixture. In the cases where the source activation patterns are unknown, such systems have to either adjust the number of outputs or to identify invalid outputs from the valid ones. Iterative separation methods have gain much attention in the community as they can flexibly decide the number of outputs, however (1) they typically rely on long-term information to determine the stopping time for the iterations, which makes them hard to operate in a causal setting; (2) they lack a ``fault tolerance'' mechanism when the estimated number of sources is different from the actual number. In this paper, we propose a simple training method, the auxiliary autoencoding permutation invariant training (A2PIT), to alleviate the two issues. A2PIT assumes a fixed number of outputs and uses auxiliary autoencoding loss to force the invalid outputs to be the copies of the input mixture, and detects invalid outputs in a fully unsupervised way during inference phase. Experiment results show that A2PIT is able to improve the separation performance across various numbers of speakers and effectively detect the number of speakers in a mixture.

%% file: introduction.tex
Many recent source separation systems assume that the number of active sources in a mixture is known in advance during both training and inference phases \cite{hershey2016deep, isik2016single, yu2017permutation, kolbaek2017multitalker, chen2017deep, luo2017deep, wang2018alternative, luo2018tasnet, luo2019conv, wang2019pitch, liu2019divide, le2019phasebook, kavalerov2019universal, luo2019dual}. Such assumption can be valid when there is additional information, such as visual cue or source locations \cite{yoshioka2019advances, grondin2019multiple}, however for a general blind source separation system it is typically not straightforward to obtain such information, especially in inference phase. In problems such as the separation of shorter streams or chunks in a long mixture, e.g. real-world conversations or music recordings, the number of active sources can vary from chunk to chunk. The estimation of the number of valid sources in a mixture is thus an important problem towards the successful deployment of separation systems into such applications.

Various methods have been proposed to tackle the problem of separating varying numbers of sources. A most simple way is to assume a maximum number of sources in a mixture, which is denoted by $N$, and let the model to always generate $N$ outputs \cite{kolbaek2017multitalker, liu2019divide}. For mixtures having $M$ sources where $M<N$, $N-M$ outputs are invalid and need to be properly designed and effectively detected. The invalid outputs are typically forced to have a significantly smaller energy than the valid outputs, and a energy threshold can then be applied to filter out those outputs. Another approach first estimates the speaker embedding for each active source with an output-length-free model, e.g. a sequence-to-sequence generative model, and then performs speaker extraction based on the embeddings \cite{shi2019ones}. A third category of methods perform separation in an iterative way, where in each iteration only one target source is separated from the residual mixture \cite{shi2018listen, kinoshita2018listening, von2019all, takahashi2019recursive}. The iteration stops when there is no source left, and the stop time can be determined by either an energy threshold or another trained discriminator. It has been shown that under various circumstances, the number of sources in the mixture can be effectively estimated and the separation performance can be guaranteed.
 
On the other hand, there are various drawbacks in each category of the existing methods. For the fixed-output-number method, the training targets for the invalid outputs are typically low- or zero-energy signals. However, such targets cannot be jointly used with energy-invariant training objectives, such as scale-invariant signal-to-distortion ratio (SI-SDR) \cite{le2019sdr}, which has proven to be a better training objective in many scenarios \cite{kolbaek2020loss}. Moreover, the detection of invalid outputs typically relies on a pre-defined energy threshold, which may cause trouble when the mixture also has a very low energy. For the speaker extraction method, the speaker embeddings are typically estimated at utterance-level and require a long enough context, which makes the method hard to apply in online or causal systems. For methods that utilize additional target speaker enrollments for speaker embedding extraction, the generalization ability on unseen speakers is also limited. For the iterative method, the run-time complexity linearly increases as the number of sources increases, and stop time detection is typically performed at utterance-level as well. When there is noise in the mixture, it is also unclear in which iteration should the noise be cancelled. Moreover, none of the methods have a ``fault tolerance'' mechanism when the estimated number of sources is different than the actual number. What should the model append to the output if it estimates fewer sources than the actual case? How should the model delete invalid outputs if it generates more? How can such decision process or control flow be effectively incorporated into the training of the model? These questions are important for a practical and robust system.
 
In this paper, we propose a simple training method based on the fixed-output assumption by designing proper training targets for the invalid outputs. We adopt the fixed-output-number assumption as in real-world conversations such as meeting scenarios, the maximum number of simultaneously active speakers is almost always fewer than three \cite{ccetin2006analysis, yoshioka2019advances}, thus a maximum number of speakers can typically be pre-assumed. Instead of using low-energy auxiliary targets for invalid outputs, we use the mixture itself as auxiliary targets to force the invalid outputs to perform autoencoding. With the permutation invariant training (PIT) framework \cite{yu2017permutation} for speech separation, we refer to it as the auxiliary autoencoding permutation invariant training (A2PIT). A2PIT not only allows the model to perform valid output detection in a self-supervised way without additional modules, but also achieves ``fault tolerance'' by the \textit{``do nothing is better than do wrong things''} principle. As the mixture itself can be treated as the output of a null separation model, i.e. perform no separation at all, the auxiliary targets force the model to generate outputs not worse than doing nothing. Moreover, the detection of invalid outputs in A2PIT can be done at frame-level based on the similarity between the outputs and the mixture, which makes it possible to perform single-pass separation and valid source detection in real-time.

The rest of the paper is organized as follows. Section~\ref{sec:A2PIT} first makes a quick overview on the PIT framework and then introduces the proposed A2PIT method. Section~\ref{sec:exp} provides the experiment configurations and discusses the results. Section~\ref{sec:conclusion} concludes the paper.

%% file: A2PIT.tex
\subsection{Permutation Invariant Training}

Permutation Invariant Training (PIT) is currently the most widely used training method for speech separation systems. PIT aims at solving the \textit{output permutation problem} in supervised learning setting, where the correct label permutation of the training targets is unknown with respect to the model outputs. Unlike methods that explicitly use the label permutation information inside the model \cite{hershey2016deep, luo2017speaker}, PIT calculates the loss between the outputs and all possible permutations of the targets, and select the one that corresponds to the minimum loss for back-propagation.

Models using PIT for training often have a fixed number of outputs, which we denote the number as $N$. For the problem of separating varying numbers of sources where the actual number of sources are $M\leq N$, $N-M$ auxiliary targets need to be properly designed. A typical way is to use low-energy random Gaussian noise as targets and detect invalid outputs by using a simple energy threshold \cite{kolbaek2017multitalker}, and it has shown that in certain datasets this energy-based method can achieve reasonable performance.

\subsection{Auxiliary Autoencoding for Invalid Outputs}

There are two main issues in the energy-based method for invalid output detection. First, it cannot be jointly used with energy-invariant objective functions like SI-SDR. Second, once the detection of invalid speakers fails and the noise signals are selected as the targets, the outputs can be completely uncorrelated with any of the targets, which is unpreferred for applications that require high perceptual quality or low distortion. We define this as the problem of lacking ``fault tolerance'' mechanism for unsuccessful separation.

To allow the models to use any objective functions and to have such ``fault tolerance'' ability, we select the mixture signal itself as the auxiliary targets instead of random noise signals. For mixtures with $N$ outputs and $M<N$ targets, $N-M$ mixture signals are appended to the targets and PIT is applied to find the best output permutation with respect to the targets. The A2PIT loss with the best permutation then becomes:
\begin{align}
    \mathcal{L}_{obj} = \mathcal{L}_{sep} + \mathcal{L}_{AE}
\end{align}
where $\mathcal{L}_{sep} \in \mathbb{R}$ is the loss for the valid outputs and $\mathcal{L}_{AE} \in \mathbb{R}$ is the auxiliary autoencoding loss for the invalid outputs with the input mixture as targets. As autoencoding is in general a much simpler task than separation, proper gradient balancing method should be applied on the two loss terms for successful training. Recall that SI-SDR is defined as:
\begin{align}
    \text{SI-SDR}(\vec{x}, \hat{\vec{x}}) = 10\,\text{log}_{10} \frac{||\alpha\vec{x}||_2^2}{||\hat{\vec{x}} - \alpha\vec{x}||_2^2}
\end{align}
where $\alpha = \hat{\vec{x}}\vec{x}^\top / \vec{x}\vec{x}^\top$ corresponds to the optimal rescaling factor towards the estimated signal. Let $a \triangleq \vec{x}\vec{x}^\top$, $b \triangleq \hat{\vec{x}}\vec{x}^\top$ and $c \triangleq \hat{\vec{x}}\hat{\vec{x}}^\top$, we can rewrite the definition as:
\begin{align}
\begin{split}
    \text{SI-SDR}(\vec{x}, \hat{\vec{x}}) &= 10\,\text{log}_{10}\left(\frac{b^2 / a}{c-2b^2 / a + b^2 / a}\right) \\
    &= 10\,\text{log}_{10}\left(\frac{1}{ac / b^2-1}\right) \\
    &\triangleq 10\,\text{log}_{10}\left(\frac{c(\vec{x}, \hat{\vec{x}})^2}{1-c(\vec{x}, \hat{\vec{x}})^2}\right)
\label{eqn:SI-SDR}
\end{split}
\end{align}
where $c(\vec{x}, \hat{\vec{x}}) \triangleq b/\sqrt{ac} = \hat{\vec{x}}\vec{x}^\top/\sqrt{(\vec{x}\vec{x}^\top)(\hat{\vec{x}}\hat{\vec{x}}^\top)}$ is the cosine similarity between $\vec{x}$ and $\hat{\vec{x}}$. The scale-invariance behavior of SI-SDR can be easily observed by the nature of cosine similarity, and $\text{SI-SDR}(\vec{x}, \hat{\vec{x}}) \rightarrow +\infty$ as $\left|c(\vec{x}, \hat{\vec{x}})\right| \rightarrow 1$. It's easy to see that the second term in $\left|\partial\, \text{SI-SDR}(\vec{x}, \hat{\vec{x}}) / \partial\, c(\vec{x}, \hat{\vec{x}})\right|$ approaches infinity as $\left|c(\vec{x}, \hat{\vec{x}})\right|$ approaches 1. Using it for $\mathcal{L}_{AE}$ may let the system to easily collapse to a local minimum which have very high performance on the auxiliary autoencoding term while fail to separate the sources. Based on this concern, we propose an $\alpha$-skewed SI-SDR ($\alpha$-SI-SDR):
\begin{align}
    \text{$\alpha$-SI-SDR}(\vec{x}, \hat{\vec{x}}) \triangleq 10\,\text{log}_{10}\left(\frac{c(\vec{x}, \hat{\vec{x}})^2}{1+\alpha-c(\vec{x}, \hat{\vec{x}})^2}\right)
\label{eqn:a-SI-SDR}
\end{align}
where the scale of the gradient with respect to the cosine similarity term is controlled by $\alpha \geq 0$, and $\alpha=0$ corresponds to the standard SI-SDR. For multiple-speaker utterances, we empirically set $\alpha=0.3$ for $\mathcal{L}_{AE}$ and $\alpha=0$ for $\mathcal{L}_{sep}$. For single-speaker utterances, the training target for separation is equivalent (when there is no noise) or very close (when there is noise) to the input mixture. In this case, we also set $\alpha=0.3$ for $\mathcal{L}_{sep}$.

\subsection{Detection of invalid outputs}
\label{sec:select}

During inference phase, the detection of invalid outputs can be performed by calculating the similarity, e.g. SI-SDR score, between all outputs and the input mixture, and a threshold calculated from the training set can be used for the decision. For the ``fault tolerance'' mechanism, the following method is applied for selecting the valid outputs:
\begin{enumerate}
    \item If the estimated number of outputs $K$ is smaller than the actual number $M$, $M-K$ additional outputs are randomly selected from the $N-K$ remaining outputs.
    \item If the estimated number of outputs $K$ is larger than the actual number $M$, $M$ outputs are randomly selected from the $K$ outputs.
\end{enumerate}

Another benefit for A2PIT is that it also allows frame-level detection of the invalid outputs for causal applications. Frame-level detection calculates accumulated similarity starting from the first frame of the outputs, and is able to dynamically change the selected valid outputs as the similarity scores become more reliable. For streaming-based applications that require a real-time playback of the separation outputs, e.g. hearable devices, the change of the output tracks can also be easily done by switching the outputs at frame-level. We leave it as a future task and focus on utterance-level detection in Section~\ref{sec:threshold}.

%% file: experiments.tex
\begin{figure*}[!htp]
\caption{Histograms of autoencoding SI-SDR (decibel scale) in different experiment configurations.}
\small
\centering
\includegraphics[width=2\columnwidth]{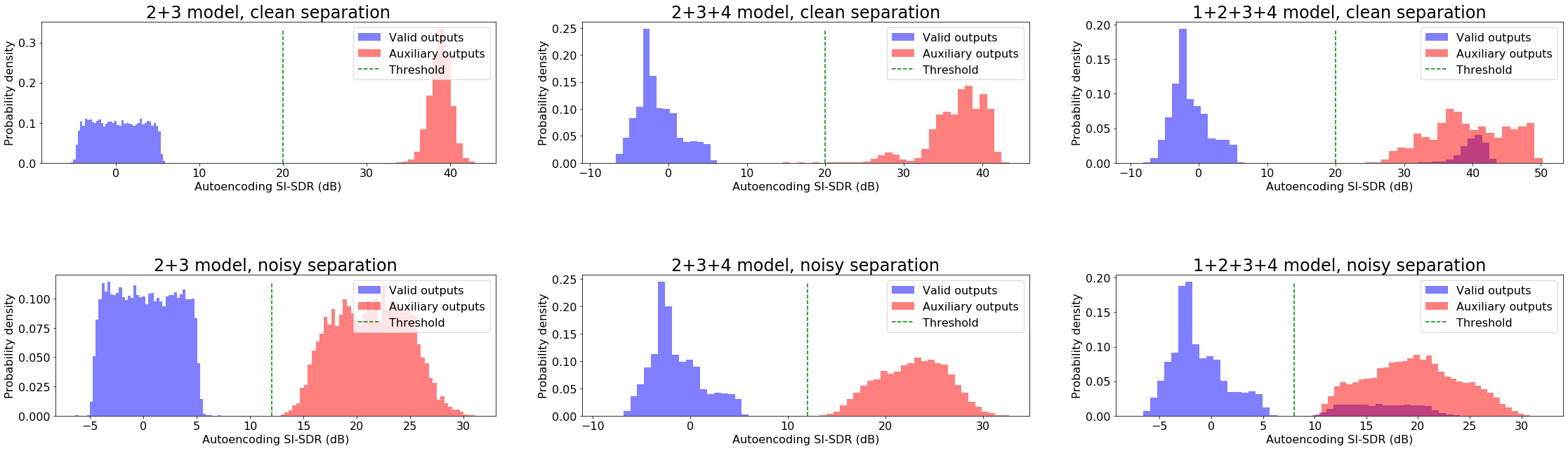}
\label{fig:hist}
\end{figure*}

\subsection{Dataset}

We simulate a single-channel noisy speech separation dataset with the Librispeech dataset \cite{panayotov2015librispeech}. 40 hours of training data (), 20 hours of validation data, and 12 hours of test data are generated from the 100-hour training set, development set, and test set, respectively. The number of speakers are evenly sampled between 1 and 4 to make sure the dataset is balanced to the varying numbers of speakers. All utterances are 6-second long with a sample rate of 16k Hz. For utterances with more than one speaker, an overlap ratio between all the speakers is uniformly sampled between 0\% and 100\% and the speech signals are shifted accordingly. The speech signals are then rescaled to a random absolute energy between -2.5 and 2.5 dB. A noise signal is randomly selected from the 100 Nonspeech Corpus \cite{web100nonspeech}, and is repeated if its length is less than 6 seconds. The noise signal is then rescaled to a random absolute energy between -20 and -10 dB. We use both the clean and noisy mixtures to report the performance of A2PIT in the two scenarios.

\subsection{Model configurations}

We adopt the time-domain audio separation network (TasNet) with dual-path RNN (DPRNN) \cite{luo2019dual} for all experiments. We use the same hyperparameter settings as in \cite{luo2019dual} for the 2 ms window configuration, with the only difference that we use 3 instead of 6 DPRNN blocks. The total number of parameters is thus 1.3M. The baseline model uses the standard SI-SDR as the training objective, and all other models use the proposed A2PIT together with $\alpha$-SI-SDR proposed in Section~\ref{sec:A2PIT}. All models are trained for a maximum of 100 epochs with the Adam optimizer \cite{kingma2014adam}. The initial learning rate is $1e-3$ and is decayed by a factor of 0.98 for every two epochs. No other regularizers or training tricks are applied.

We train the DPRNN-TasNet for each of the speaker count configurations as the baseline models. These results represents how well the models can achieve when the number of speakers is known and a specific model is trained on such mixtures. For separating varying numbers of sources, we train DPRNN-TasNet models on three configurations:
\begin{enumerate}
    \item 2+3 speakers: the 2 and 3 speaker mixtures are used for both training and evaluation, and the number of outputs $N$ is set to 3. This is to mimic the behavior under certain cases when the maximum number of active sources is bounded by 3 (e.g. meeting scenarios). We denote it as the \textit{2+3 model}.
    \item 2+3+4 speakers: the 2, 3 and 4 speaker mixtures are used for both training and evaluation, and the number of outputs $N$ is set to 4. This is to increase the difficulty of both the separation and speaker count. We denote it as the \textit{2+3+4 model}.
    \item 1+2+3+4 speakers: all training and evaluation datasets are used. We denote it as the \textit{1+2+3+4 model}.
\end{enumerate}

Each configuration contains both the clean and noisy scenarios, which results in a total of 6 different configurations.

\subsection{Evaluation metrics}

We evaluate the separation performance by the SI-SDR improvement (SI-SDRi) with respect to the unprocessed mixture. To evaluate the accuracy for speaker number detection, we report the confusion matrix of the predicted and oracle numbers of speakers in the test set.

\subsection{Results and discussions}

\subsubsection{Determine the similarity threshold through the training set}
\label{sec:threshold}
The similarity threshold described in Section~\ref{sec:select} needs to be determined through the training set. Figure~\ref{fig:hist} shows the autoencoding SI-SDR, i.e. the SI-SDR between the outputs and the input mixture, for different configurations. We can see that in the clean separation task, most of the auxiliary outputs have a significantly higher autoencoding SI-SDR than the valid outputs. The high SI-SDR utterances in the \textit{1+2+3+4 model} are mainly due to the single-speaker samples. This allows us to draw a clear boundary to distinguish them. We empirically set the threshold for all models for clean separation tasks to be 20, i.e. outputs with autoencoding SI-SDR higher than 20 dB will be treated as invalid outputs, and set the thresholds for the 2+3 model, 2+3+4 model, 1+2+3+4 model for the noisy separation tasks to be 12, 12, and 8, respectively.

\subsubsection{Accuracy of speaker counting}

\begin{table}[!htp]
\caption{Confusion matrix for speaker counting for models trained for clean separation task.}
\small
\centering
\begin{tabular}{c|c|c|c|c|c}
\multirow{2}{*}{Model} & \multirow{2}{*}{Prediction} & \multicolumn{4}{c}{Oracle} \\
\cline{3-6}
& & 1 spk & 2 spk & 3 spk & 4 spk \\
\thline
\multirow{2}{*}{\makecell{\textit{2+3}\\\textit{model}}} & 2 spk & -- & 1712 & 5 & -- \\
& 3 spk & -- & 88 & 1795 & -- \\
\hline
\multirow{3}{*}{\makecell{\textit{2+3+4}\\\textit{model}}} & 2 spk & -- & 1718 & 10 & 0 \\
& 3 spk & -- & 82 & 1435 & 26 \\
& 4 spk & -- & 0 & 355 & 1774 \\
\hline
\multirow{4}{*}{\makecell{\textit{1+2+3+4}\\\textit{model}}} & 1 spk & 2 & 0 & 0 & 0 \\
& 2 spk & 5 & 1746 & 13 & 2 \\
& 3 spk & 0 & 62 & 1454 & 44  \\
& 4 spk & 0 & 0 & 333 & 1756 \\
\thline
\end{tabular}
\label{tab:count-clean}
\end{table}

\begin{table}[!htp]
\caption{Confusion matrix for speaker counting for models trained for noisy separation task.}
\small
\centering
\begin{tabular}{c|c|c|c|c|c}
\multirow{2}{*}{Model} & \multirow{2}{*}{Prediction} & \multicolumn{4}{c}{Oracle} \\
\cline{3-6}
& & 1 spk & 2 spk & 3 spk & 4 spk \\
\thline
\multirow{2}{*}{\makecell{\textit{2+3}\\\textit{model}}} & 2 spk & -- & 1716 & 26 & -- \\
& 3 spk & -- & 83 & 1774 & -- \\
\hline
\multirow{3}{*}{\makecell{\textit{2+3+4}\\\textit{model}}} & 2 spk & -- & 1711 & 16 & 0 \\
& 3 spk & -- & 87 & 1530 & 87 \\
& 4 spk & -- & 1 & 254 & 1713 \\
\hline
\multirow{4}{*}{\makecell{\textit{1+2+3+4}\\\textit{model}}} & 1 spk & 31 & 4 & 0 & 0 \\
& 2 spk & 5 & 1670 & 8 & 0 \\
& 3 spk & 0 & 125 & 1485 & 27  \\
& 4 spk & 0 & 0 & 307 & 1773 \\
\thline
\end{tabular}
\label{tab:count-noisy}
\end{table}

Table~\ref{tab:count-clean} and~\ref{tab:count-noisy} show the confusion matrices for all 6 configurations. Note that each of the speaker number has a test set of 1800 utterances. We first notice that for the \textit{2+3 model}, the prediction of speaker count can be done with a very high accuracy in both clean and noisy separation tasks. For the \textit{2+3+4 model}, the detection of 3 speaker mixtures is worse than that of both 2 and 4 speaker mixtures, and the error mostly comes from the misclassification into 4 speaker mixtures. For the \textit{1+2+3+4 model}, we find that the detection of the 1 speaker mixtures almost always fail (detects no speakers in the mixture). With the autoencoding threshold, the model predicts no valid outputs for most of the times. This is somehow expected as in the clean separation task, the mixture itself is equivalent to the separated output, and in the noisy separation task, the separated output may still have very high similarity score with respect to the mixture because of our high SNR configuration. For tasks such as automatic speech recognition, this will not be an issue as the acoustic models are typically noise robust, while for tasks that require perceptual quality, the outputs need to be further evaluated. Beyond the 1 speaker mixtures, the accuracy for speaker counting for other cases remains high.

Another interesting observation is that the models occasionally predict zero speakers (e.g. 2-speaker utterances in all models for noisy separation). This can only happen when the autoencoding SI-SDR of all outputs are larger than the pre-defined threshold. It indicates that in certain utterances the separation may completely fail and the model converges to always perform autoencoding. A better solution to this issue is left for future works.

\subsubsection{Performance of speech separation}

\begin{table}[!htp]
\caption{Separation performance of various configurations on the clean separation task. SI-SDR is reported for one speaker utterances in decibel scale, and SI-SDRi is reported for the rest in decibel scale.}
\small
\centering
\begin{tabular}{c|c|c|c|c|c}
\multirow{2}{*}{Model} & \multirow{2}{*}{\makecell{Output\\selection}} & SI-SDR & \multicolumn{3}{c}{SI-SDRi} \\
\cline{3-6}
& & 1 spk & 2 spk & 3 spk & 4 spk \\
\thline
\makecell{Baseline} & Oracle & \textbf{64.8} & 11.5 & 8.0 & 5.7 \\
\hline
\multirow{2}{*}{\makecell{\textit{2+3}\\\textit{model}}} & Oracle & -- & \textbf{12.0} & 8.8 & -- \\
& Predicted & -- & 11.6 & 8.7 & -- \\
\hline
\multirow{2}{*}{\makecell{\textit{2+3+4}\\\textit{model}}} & Oracle & -- & 11.8 & \textbf{9.1} & 7.1 \\
& Predicted & -- & 11.7 & 8.1 & 7.1 \\
\hline
\multirow{2}{*}{\makecell{\textit{1+2+3+4}\\\textit{model}}} & Oracle & 39.8 & 11.9 & \textbf{9.1} & \textbf{7.2} \\
& Predicted & 44.2 & 11.8 & 8.5 & \textbf{7.2} \\
\end{tabular}
\label{tab:sep-clean}
\end{table}

\begin{table}[!htp]
\caption{Separation performance of various configurations on the noisy separation task. SI-SDRi is reported in decibel scale.}
\small
\centering
\begin{tabular}{c|c|c|c|c|c}
\multirow{2}{*}{Model} & \multirow{2}{*}{\makecell{Output\\selection}} & \multicolumn{4}{c}{SI-SDRi} \\
\cline{3-6}
& & 1 spk & 2 spk & 3 spk & 4 spk \\
\thline
\makecell{Baseline} & Oracle & \textbf{6.9} & 10.8 & 7.5 & 5.4 \\
\hline
\multirow{2}{*}{\makecell{\textit{2+3}\\\textit{model}}} & Oracle & -- & \textbf{11.2} & 8.7 & -- \\
& Predicted & -- & \textbf{11.2} & 8.7 & -- \\
\hline
\multirow{2}{*}{\makecell{\textit{2+3+4}\\\textit{model}}} & Oracle & -- & 11.1 & \textbf{8.8} & \textbf{7.0} \\
& Predicted & -- & 10.8 & 8.2 & 6.9 \\
\hline
\multirow{2}{*}{\makecell{\textit{1+2+3+4}\\\textit{model}}} & Oracle & 4.8 & 11.1 & \textbf{8.8} & 6.9 \\
& Predicted & 4.2 & 11.0 & 8.4 & 6.9 \\
\end{tabular}
\label{tab:sep-noisy}
\end{table}

Table~\ref{tab:sep-clean} and~\ref{tab:sep-noisy} provide the separation performance on the clean and noisy separation tasks, respectively. For the one speaker utterances in the clean separation task, SI-SDR instead of SI-SDRi is reported as the input is already the clean target itself. We observe that A2PIT can almost always improve the separation performance on all configurations with both clean and noisy data, and the gains for 3 and 4 speaker cases are significant. We can conclude from the results that A2PIT is able to achieve on par or better overall separation performance on both clean and noisy separation tasks. Even with predicted output selection, the fault tolerance ability introduced by A2PIT allows the model to control the performance degradation. These results confirms the effectiveness of A2PIT.

%% file: conclusion.tex
In this paper, we proposed a simple method for separating varying numbers of speakers in a mixture with ``fault tolerance'' ability, which we referred to as the auxiliary autoencoding permutation invariant training (A2PIT). A2PIT assumed a fixed number of outputs $N$ and appended mixture signals to the training targets of the utterances whose number of valid outputs $M$ was smaller than $N$. Fault tolerance was achieved by treating the auxiliary outputs as the outputs of a ``null'' separation which directly passed the input to the output. We call this the ``do nothing is better than do wrong things'' principle. During inference time, a similarity threshold between the mixture and the outputs was used to determine valid outputs in a fully unsupervised way. Experiment results showed that A2PIT was able to effectively perform speaker count in various scenarios, and maintained on par or better separation performance than baseline systems trained for specific datasets with both oracle and predicted speaker count.

%% file: main.bbl
\begin{thebibliography}{10}
\providecommand{\url}[1]{#1}
\csname url@samestyle\endcsname
\providecommand{\newblock}{\relax}
\providecommand{\bibinfo}[2]{#2}
\providecommand{\BIBentrySTDinterwordspacing}{\spaceskip=0pt\relax}
\providecommand{\BIBentryALTinterwordstretchfactor}{4}
\providecommand{\BIBentryALTinterwordspacing}{\spaceskip=\fontdimen2\font plus
\BIBentryALTinterwordstretchfactor\fontdimen3\font minus
  \fontdimen4\font\relax}
\providecommand{\BIBforeignlanguage}[2]{{%
\expandafter\ifx\csname l@#1\endcsname\relax
\typeout{** WARNING: IEEEtran.bst: No hyphenation pattern has been}%
\typeout{** loaded for the language `#1'. Using the pattern for}%
\typeout{** the default language instead.}%
\else
\language=\csname l@#1\endcsname
\fi
#2}}
\providecommand{\BIBdecl}{\relax}
\BIBdecl

\bibitem{hershey2016deep}
J.~R. Hershey, Z.~Chen, J.~Le~Roux, and S.~Watanabe, ``Deep clustering:
  Discriminative embeddings for segmentation and separation,'' in
  \emph{Acoustics, Speech and Signal Processing (ICASSP), 2016 IEEE
  International Conference on}.\hskip 1em plus 0.5em minus 0.4em\relax IEEE,
  2016, pp. 31--35.

\bibitem{isik2016single}
Y.~Isik, J.~Le~Roux, Z.~Chen, S.~Watanabe, and J.~R. Hershey, ``Single-channel
  multi-speaker separation using deep clustering,'' \emph{Interspeech 2016},
  pp. 545--549, 2016.

\bibitem{yu2017permutation}
D.~Yu, M.~Kolb{\ae}k, Z.-H. Tan, and J.~Jensen, ``Permutation invariant
  training of deep models for speaker-independent multi-talker speech
  separation,'' in \emph{Acoustics, Speech and Signal Processing (ICASSP), 2017
  IEEE International Conference on}.\hskip 1em plus 0.5em minus 0.4em\relax
  IEEE, 2017, pp. 241--245.

\bibitem{kolbaek2017multitalker}
M.~Kolb{\ae}k, D.~Yu, Z.-H. Tan, and J.~Jensen, ``Multitalker speech separation
  with utterance-level permutation invariant training of deep recurrent neural
  networks,'' \emph{IEEE/ACM Transactions on Audio, Speech, and Language
  Processing (TASLP)}, vol.~25, no.~10, pp. 1901--1913, 2017.

\bibitem{chen2017deep}
Z.~Chen, Y.~Luo, and N.~Mesgarani, ``Deep attractor network for
  single-microphone speaker separation,'' in \emph{Acoustics, Speech and Signal
  Processing (ICASSP), 2017 IEEE International Conference on}.\hskip 1em plus
  0.5em minus 0.4em\relax IEEE, 2017, pp. 246--250.

\bibitem{luo2017deep}
Y.~Luo, Z.~Chen, J.~R. Hershey, J.~Le~Roux, and N.~Mesgarani, ``Deep clustering
  and conventional networks for music separation: Stronger together,'' in
  \emph{Acoustics, Speech and Signal Processing (ICASSP), 2017 IEEE
  International Conference on}.\hskip 1em plus 0.5em minus 0.4em\relax IEEE,
  2017, pp. 61--65.

\bibitem{wang2018alternative}
Z.-Q. Wang, J.~Le~Roux, and J.~R. Hershey, ``Alternative objective functions
  for deep clustering,'' in \emph{Acoustics, Speech and Signal Processing
  (ICASSP), 2018 IEEE International Conference on}, 2018.

\bibitem{luo2018tasnet}
Y.~Luo and N.~Mesgarani, ``Tas{N}et: time-domain audio separation network for
  real-time, single-channel speech separation,'' in \emph{Acoustics, Speech and
  Signal Processing (ICASSP), 2018 IEEE International Conference on}.\hskip 1em
  plus 0.5em minus 0.4em\relax IEEE, 2018.

\bibitem{luo2019conv}
------, ``Conv-{T}as{N}et: Surpassing ideal time--frequency magnitude masking
  for speech separation,'' \emph{IEEE/ACM Transactions on Audio, Speech, and
  Language Processing (TASLP)}, vol.~27, no.~8, pp. 1256--1266, 2019.

\bibitem{wang2019pitch}
K.~Wang, F.~Soong, and L.~Xie, ``A pitch-aware approach to single-channel
  speech separation,'' in \emph{Acoustics, Speech and Signal Processing
  (ICASSP), 2019 IEEE International Conference on}.\hskip 1em plus 0.5em minus
  0.4em\relax IEEE, 2019, pp. 296--300.

\bibitem{liu2019divide}
Y.~Liu and D.~Wang, ``Divide and conquer: A deep casa approach to
  talker-independent monaural speaker separation,'' \emph{IEEE/ACM Transactions
  on Audio, Speech, and Language Processing (TASLP)}, vol.~27, no.~12, pp.
  2092--2102, 2019.

\bibitem{le2019phasebook}
J.~Le~Roux, G.~Wichern, S.~Watanabe, A.~Sarroff, and J.~R. Hershey, ``The
  phasebook: Building complex masks via discrete representations for source
  separation,'' in \emph{Acoustics, Speech and Signal Processing (ICASSP), 2019
  IEEE International Conference on}.\hskip 1em plus 0.5em minus 0.4em\relax
  IEEE, 2019, pp. 66--70.

\bibitem{kavalerov2019universal}
I.~Kavalerov, S.~Wisdom, H.~Erdogan, B.~Patton, K.~Wilson, J.~Le~Roux, and
  J.~R. Hershey, ``Universal sound separation,'' in \emph{2019 IEEE Workshop on
  Applications of Signal Processing to Audio and Acoustics (WASPAA)}.\hskip 1em
  plus 0.5em minus 0.4em\relax IEEE, 2019, pp. 175--179.

\bibitem{luo2019dual}
Y.~Luo, Z.~Chen, and T.~Yoshioka, ``Dual-path {RNN}: efficient long sequence
  modeling for time-domain single-channel speech separation,'' \emph{arXiv
  preprint arXiv:1910.06379}, 2019.

\bibitem{yoshioka2019advances}
T.~Yoshioka, I.~Abramovski, C.~Aksoylar, Z.~Chen, M.~David, D.~Dimitriadis,
  Y.~Gong, I.~Gurvich, X.~Huang, Y.~Huang \emph{et~al.}, ``Advances in online
  audio-visual meeting transcription,'' \emph{arXiv preprint arXiv:1912.04979},
  2019.

\bibitem{grondin2019multiple}
F.~Grondin and J.~Glass, ``Multiple sound source localization with svd-phat,''
  \emph{Interspeech 2019}, pp. 2698--2702, 2019.

\bibitem{shi2019ones}
J.~Shi, J.~Xu, and B.~Xu, ``Which ones are speaking? speaker-inferred model for
  multi-talker speech separation,'' \emph{Interspeech 2019}, pp. 4609--4613,
  2019.

\bibitem{shi2018listen}
J.~Shi, J.~Xu, G.~Liu, and B.~Xu, ``Listen, think and listen again: capturing
  top-down auditory attention for speaker-independent speech separation,'' in
  \emph{Proceedings of the 27th International Joint Conference on Artificial
  Intelligence}.\hskip 1em plus 0.5em minus 0.4em\relax AAAI Press, 2018, pp.
  4353--4360.

\bibitem{kinoshita2018listening}
K.~Kinoshita, L.~Drude, M.~Delcroix, and T.~Nakatani, ``Listening to each
  speaker one by one with recurrent selective hearing networks,'' in
  \emph{Acoustics, Speech and Signal Processing (ICASSP), 2018 IEEE
  International Conference on}.\hskip 1em plus 0.5em minus 0.4em\relax IEEE,
  2018, pp. 5064--5068.

\bibitem{von2019all}
T.~von Neumann, K.~Kinoshita, M.~Delcroix, S.~Araki, T.~Nakatani, and
  R.~Haeb-Umbach, ``All-neural online source separation, counting, and
  diarization for meeting analysis,'' in \emph{Acoustics, Speech and Signal
  Processing (ICASSP), 2019 IEEE International Conference on}.\hskip 1em plus
  0.5em minus 0.4em\relax IEEE, 2019, pp. 91--95.

\bibitem{takahashi2019recursive}
N.~Takahashi, S.~Parthasaarathy, N.~Goswami, and Y.~Mitsufuji, ``Recursive
  speech separation for unknown number of speakers,'' \emph{Interspeech 2019},
  pp. 1348--1352, 2019.

\bibitem{le2019sdr}
J.~Le~Roux, S.~Wisdom, H.~Erdogan, and J.~R. Hershey, ``{SDR}--half-baked or
  well done?'' in \emph{Acoustics, Speech and Signal Processing (ICASSP), 2019
  IEEE International Conference on}.\hskip 1em plus 0.5em minus 0.4em\relax
  IEEE, 2019, pp. 626--630.

\bibitem{kolbaek2020loss}
M.~Kolb{\ae}k, Z.-H. Tan, S.~H. Jensen, and J.~Jensen, ``On loss functions for
  supervised monaural time-domain speech enhancement,'' \emph{IEEE/ACM
  Transactions on Audio, Speech, and Language Processing (TASLP)}, 2020.

\bibitem{ccetin2006analysis}
{\"O}.~{\c{C}}etin and E.~Shriberg, ``Analysis of overlaps in meetings by
  dialog factors, hot spots, speakers, and collection site: insights for
  automatic speech recognition,'' in \emph{Ninth International Conference on
  Spoken Language Processing}, 2006.

\bibitem{luo2017speaker}
\BIBentryALTinterwordspacing
Y.~Luo, Z.~Chen, and N.~Mesgarani, ``Speaker-independent speech separation with
  deep attractor network,'' \emph{IEEE/ACM Transactions on Audio, Speech, and
  Language Processing (TASLP)}, vol.~26, no.~4, pp. 787--796, 2018. [Online].
  Available: \url{http://dx.doi.org/10.1109/TASLP.2018.2795749}
\BIBentrySTDinterwordspacing

\bibitem{panayotov2015librispeech}
V.~Panayotov, G.~Chen, D.~Povey, and S.~Khudanpur, ``Librispeech: an {ASR}
  corpus based on public domain audio books,'' in \emph{Acoustics, Speech and
  Signal Processing (ICASSP), 2015 IEEE International Conference on}.\hskip 1em
  plus 0.5em minus 0.4em\relax IEEE, 2015, pp. 5206--5210.

\bibitem{web100nonspeech}
G.~Hu, ``100 {N}onspeech {S}ounds,'' {\small
  \url{http://web.cse.ohio-state.edu/pnl/corpus/HuNonspeech/HuCorpus.html}}.

\bibitem{kingma2014adam}
D.~Kingma and J.~Ba, ``Adam: A method for stochastic optimization,''
  \emph{arXiv preprint arXiv:1412.6980}, 2014.

\end{thebibliography}
